# Minimum Energy Information Fusion in Sensor Networks

George Chapline
Lawrence Livermore National Laboratory

Abstract

In this paper we consider how to organize the sharing of information in a distributed network of sensors and data processors so as to provide explanations for sensor readings with minimal expenditure of energy. We point out that the Minimum Description Length principle provides an approach to information fusion that is more naturally suited to energy minimization than traditional Bayesian approaches. In addition we show that for networks consisting of a large number of identical sensors Kohonen self-organization provides an exact solution to the problem of combing the sensor outputs into minimal description length explanations.

## 1. Introduction

One of the grand challenges of cognitive science is to understand how, at least in principle, a network of sensors and simple data processors might be configured to "understand" what is going on its environment. In general forming perceptions from sensor outputs will require a network of sensors because noise or insufficient selectivity may not allow individual sensors to provide unambiguos signals regarding the environment. It should be kept in mind in this connection that increasing the sensitivity of an individual detector will not lead to an increase in the signal to noise ratio for the signatures of interest unless some scheme for background subtraction is available. The upshot is that even in networks where the individual detectors are very sensitive, it will in general be desirable to correlate or "fuse" the signals from different kinds of sensors or sensors in different spatial locations.

When one is considering the problem of combing information from different sensors it is tempting to use Bayesian probabilistic reasoning [1] or its Demster-Shafer generalization [2]. One of the attractive features of a Bayesian approach to information fusion is its adaptability to incremental computational schemes [1], which allow one to pool the evidence from different sensors hierarchically using a tree-like network. In particular each node of a Bayesian data fusion tree combines the conditional probabilities for the units which proceed it in some ordering to form a new set of conditional probabilities. These Bayesian networks often incorporate unobserved latent variables known as hidden variables, and such networks have been successfully used for some quite difficult real world pattern recognition problems such as speech recognition. Bayesian-type networks are also attractive for combining the outputs of neural network classifiers [2]. On the other hand when applied to the problem of information fusion in an autonomous network of sensors and associated data processors neither Bayesian probabilistic reasoning nor the Dempster-Shafer method seem by themselves to offer any particular insights into the important problem of how to minimize overall energy usage in the network. In this paper we will argue that in contrast with Bayesian techniques the Minimum Description Length (MDL) principle [3] appears to be an ideal statistical inference methodology to use when energy usage in the network is an important constraint.

The MDL principle has been gaining in popularity as a fundamental alternative to Bayesian reasoning for statistical inference for several reasons. Two well-known problems with Bayesian reasoning are a) a priori probability distributions may not be known or even exist, and b) Bayesian methods are impractical when there are many possible explanations for a given instance of environmental data. While the Neyman-Pearson likelihood ratio test is uncontested as the best thing to use when there is just one hypothesis to test, there is no similarly canonical method when there are many approximately equally likely explanations for the environmental data. Indeed, not only does keeping track of the conditional probabilities for a possibly exponentially large number of hypotheses make hierarchical Bayesian fusion schemes difficult to implement, but choosing the single largest conditional probability to select a particular hypothesis could give the wrong answer. On the other hand the MDL principle was the inspiration for the Helmholtz machine [4], which is a promising approach to dealing with the combinatorial complexities associated with data whose explanation is ambiguous.

A third general problem with Bayesian methods is that they don't by themselves address the important question of minimizing the complexity of coded representations. A corollary of this second point is that Bayesian methods don't seem to be particularly well suited to the problem of optimizing the energy usage in an sensor network. However by focusing on the simplest possible way to explain environmental data the MDL principle appears to be very well suited to minimizing energy usage in a sensor network. In the following section we briefly review the MDL approach to pattern recognition. The basic idea here is that overall description costs are minimized when the probabilities for various explanations are related to their description costs by the Boltzmann distribution. In section 3 we show that MDL explanations for the outputs of a large number of identical sensors can obtained using Kohonen's algorithm for self-organization. In section 4 we compare the energy requirements for sensor fusion using distributed self-organization with the energy requirements for sensor fusion using a hierarchical Bayesian network.

## 2. Minimal description length approach to pattern recognition

It has been understood for some time that pattern recognition systems are in essence machines that utilize either preconceived probability distributions or empirically determined posterior probabilities to classify patterns [5]. In the ideal case where the a priori probability distribution $p(\alpha)$ for the occurrence of various classes $\alpha$ of feature vectors and probability densities $p(x|\alpha)$ for the distribution of data sets x within each class are known, then the best possible classification procedure would be to simply choose the class $\alpha$ for which the posterior probability

$$P(\alpha \mid x) = \frac{p(\alpha)p(x \mid \alpha)}{\sum_{\beta} p(\beta)p(x \mid \beta)} \qquad (1)$$

is largest. Unfortunately in the real world one is typically faced with the situation that neither the class probabilities $p(\alpha)$ nor class densities $p(x|\alpha)$ are precisely known, so that one must rely on empirical information to estimate the conditional probabilities $P(\alpha \mid x)$ needed to classify data sets. In practice this means that one must adopt a parametric model for the class probabilities and densities, and then use empirical data to fix the parameters $\theta$ of the probability model. Once values for the model parameters have been fixed, then sensory data can be classified by simply substituting values for the model probabilities $p(\alpha; \theta)$ and $p(x|\alpha; \theta)$ into equation (1).

Unfortunately determining values for the model parameters from empirical data is itself a computationally intractable problem. This means that in practice one is usually limited to using models of relatively modest complexity, and consequently one is always faced with the issue of choosing the best possible values for the model parameters. A very elegant approach to this problem was suggested some time ago by Rissanen [3], who suggested choosing a model such that the length of binary code needed to represent the model is minimized. The description length for a binary variable $s_i$ is:

$$E(s_i) = - s_i \log p_i - (1- s_i)\log(i- p_i), \qquad (2)$$

This leads us to define the description cost or "energy" of a classification $\alpha$ to be

$$E_{\alpha} = \sum_i E(s_i) + \sum_i E(x), \qquad (3)$$

where the $x$ are the variables needed to describe the input data and the $s_i$ are the variables needed to represent the interpretation of the input data. One might think that the pattern recognition algorithm should be chosen to minimize $E_{\alpha}$, but this is incorrect because it is possible [3] to devise coding schemes that take advantage of the entropy of alternative explanations for the input data. The effective cost $F(x)$ for describing a data set x with explanations $\alpha=\{s_i\}$ is

$$F(x) = \sum_\alpha \{E_\alpha P(\alpha) - (-P(\alpha)\log P(\alpha))\}. \tag{4}$$

The quantity $\sum_\alpha P_\theta(\alpha)\log[P_\theta(\alpha)/P(\alpha)]$ in the second term in equation (4) is always positive and measures of the difference in bits between the model distribution $P_\theta(\alpha)$ and the true distribution $P(\alpha)$. This distance measure, known as the Kullback- Leibler divergence, is the basis for the Maximum Likelihood estimator that is widely used by statisticians to measure how well a given set of model probabilities reproduces the empirical data [5]. As in physics $F(x)$ is minimized when the probabilities of alternative explanations are exponentially related to their costs by the canonical Boltzmann distribution:

$$P(\alpha \mid x) = \frac{e^{-E_\alpha}}{\sum_\alpha e^{-E_\alpha}}. \tag{5}$$

Thus a minimal cost recognition model should produce a probability distribution $Q(\alpha)$ that is as similar as possible to the Boltzmann distribution (5).

Of course we are still left with the problem of how to generate explanations and conditional probabilities $P(\alpha)$ that satisfy equation (5). An ingenious approach to generating explanations $\{\alpha\}$ for which the posterior probabilities $P(\alpha|x;\theta)$ are naturally represented in the canonical Boltzmann distribution form (5) was introduced in 1985 by Ackley, Hinton, and Sejnowski [6]. In this model, known as the *Boltzmann machine*, environmental data and their "explanations" are represented by configurations of binary units with activation levels $a_i = 0$ or 1. The energy function for the assembly of binary units is assumed to have the same form as that used by physicists to describe a system of interacting spins in a magnet:

$$E(\mathbf{a}) = -\frac{1}{2}\sum_{i \neq j} w_{ij} a_i a_j + \sum_i \theta_i a_i, \tag{6}$$

where $\mathbf{a} = \{a_i\}$ denotes the set of activation levels and the weight $w_{ij}$ describes the interaction strength between binary units i and j In the original version of the Boltzmann machine these interactions are assumed to be symmetric; i.e. $w_{ij} = w_{ji}$. However layered versions of the Boltzmann machine with asymmetric weights; i.e. $w_{ij} \neq w_{ji}$, are also of interest because they are equivalent to Bayesian decision networks [7]. In both the symmetric and asymmetric Boltzmann machines the probability distribution $P_\theta(\alpha)$ will be the probability distribution for the activation levels in a certain subset, referred to as the hidden units, of all binary units. The remaining binary units, referred to as the visible units, represent the environmental data x. The model parameters $\theta$ for the Boltzmann machine are the connection strengths $w_{ij}$ and biases $\theta_i$ for the binary units. These parameters are determined by minimizing the Kullbach-Leibler divergence between the probability distribution $P_\theta(\alpha)$ with the visible unit activation levels fixed and the probability distribution for classifications with the activation levels of the visible units allowed to vary freely. Used as a pattern recognition device the Boltzmann machine has the virtue that high order correlations between different instances of environmental data can be represented and used in the classification of data sets. This means that the classifications provided by the Boltzmann machine take into account more information than just the relationship between a class and its feature vectors. Unfortunately Boltzmann machines have not found many practical applications because determination of the connection strengths and biases for realistic data sets is very slow because of the necessity for repetitive Monte Carlo sampling of a joint probability distribution for the activation levels of the hidden units.

A quasi-deterministic version of the Boltzmann machine, known as the Helmholtz machine [4, 8], assumes the binary nodes are organized into layers and that there is Markov transition probability for going between layers of the form

$$p(a_i(n+1) | a(n)) = \sigma[\beta(1-2a_i(n+1))\sum_j w_{ij} a_j],  \quad (7)$$

where $\sigma(x) = 1/[1+\exp(-x)]$ and for each node of the activation level $a_i = 0$ or 1. The vector $a(n) = \{a_i(n)\}$ in equation (7) denotes the set of activation levels at layer n of the network. One can also think of the way activation levels vary from layer to layer as describing the time evolution of a system of binary units [9]. If one assumes that the activities of the binary units within a given layer are independent, then the probability of a particular explanation $\alpha$, which we identify as the "time history" $\{a(n), n>1\}$ of activations will be given by [8]:

$$Q(\alpha) = \prod_{n>1} \prod_j [p(a_j)]^{a_j} [1-p(a_j)]^{1-a_j} ;  \quad (8)$$

so that the binary units that are turned on contribute with weight $p_j(x)$ while the units that are turned off contribute with weight $1-p_j(x)$. In order to determine the recognition weights $w_{ij}$ Hinton et. al. employ a parallel "fantasy" generation network to generate a model distribution $P_\theta(\alpha)$. The weights of the fantasy generation network are chosen so as to minimize the Kullback- Leibler divergence between the model distribution and the probability values for training the recognition connection weights $w_{ij}$ using standard neural network training algorithms.

By restricting its attention to distributions of the form (8) the Helmholtz machine finesses the combinatorial problem associated many hypotheses. Therefore organizing a network of sensors and data processors as a Helmholtz machine, as was previously recommended by the author [10], might seem like a good idea. However, two aspects of the Helmholtz machine architecture seem problematical in connection with the problem of energy minimization. The first is that even though the Helmholtz machine attempts to minimize the free energy $F(x)$, by restricting attention to distributions of the form (8) it is not clear how close one can approach to the ideal Boltzmann distribution (5). A second problem is that each node in a given layer will in general be connected to every node in the previous layer. Compared with a hierarchical Bayesian network this would increase the number of communication links in a network of N total nodes by a factor on the order of N/L, where L is the number of layers. However, replacing the fully interconnected network used in the Boltzmann machine with the quasi-deterministic evolution of a string of bits does point us in the direction of the exact model for MDL information fusion described in the next section.

## 3. Self-organization approach to MDL information fusion

Let us suppose that our sensor network consists of N feature detectors, and that each feature detector can communicate with three neighboring feature detectors. The assumption of three communication links per node is made for convenience since models where the feature detectors are allowed to connect to a larger (but fixed) number of neighbors lead to similar results. Also for simplicity we will assume that the feature being looked for can be characterized by a single continuous variable w such that 0≤w≤2π; leaving for the future the more typical case where the features are characterized by a vector in a higher dimensional space. In addition we assume that every sensor is looking at the same environment. As an initial condition for the network we assign to each sensor a value $\phi$ of the feature that is randomly chosen from a probability distribution for the occurrences of various features in the environment. Now intuitively it seems clear that since in principle nearby sensors ought to have the similar outputs, a minimal description of the sensor outputs ought to involve just giving the parameters of a smooth curve for w vs. location **r** of the sensor nodes. Therefore we will guess that the data processing required for minimum description information fusion can be modeled by assuming that the maps of sensor locations into feature space are "self-organizing. If we follow Kohonen's prescription for self-organization [11], this means that the orientation of the feature detector located at r will evolve according to a rule of the form

$$w(r,t+1) = w(r,t) + h(r-s)[\phi - w(r,t)] \quad (9)$$

where h(r) is typically assumed to be a Gaussian function peaked at r=0. For our purposes the function h(r-s) can be replaced by the rule that each feature detector is connected to just three of its nearest neighbors. The location s in (9) corresponds to the feature detector whose orientation w(s) is closest to $\phi$. Thus the data fusion process is modeled as a Markov process whose states are the sets {w(r)} of possible states of the feature detectors, and where the transition probabilities are determined by probabilities of occurrence in the environment of various orientations $\phi$. In order to construct an analytical model of this evolution process it will be useful to introduce an energy functional E[w] that satisfies

$$< P(\phi)\delta w > = -grad_w E \quad (10)$$

where $\delta w = w(r,t+1) - w(r,t)$ and $P(\phi)$ is the a priori probability distribution for the orientations of the environmental stimuli. Neglecting certain mathematical subtleties, the required energy functional is [12]

$$E[w] = \frac{1}{2} \sum_{<r,s>} \sum_{\phi \in R} P(\phi) |\phi - w(r,t)|^2 \quad (11)$$

where the sum over <r,s> runs over nearest neighbor connections and R (r) is the receptive field of the feature detector located at r; i.e. the union of all environmental stimuli that are closer to w(r, t) than any other w(s,t), where s $\neq$ r.

Given an energy functional that satisfies (10) there are standard techniques that one can use to describe the stochastic evolution of the organization of our neural network. However here we will limit our interest in how the organization of feature detectors evolves with time to noting that under the influence of the random variable $\phi$(t) the system relaxes to an asymptotic state characterized by a stationary probability distribution for various final configurations of feature vectors {w(r)}. Given the existence of an energy functional satisfying (19) the statistical properties of the set {w(r)} in this stationary state can be derived from a "partition function" Z = exp[-*F* (*x*) ] which is a sum over all possible stationary state configurations weighted with the Boltzmann factor exp(-E[w]). If we assume that the stochastic evolution of network is governed by an energy functional of the form (11) then this partition function has the form [13]:

$$Z = \sum_L \kappa^F \prod_{i=1}^{F} \int_o^{2\pi} dw(r_i) e^{-\frac{K}{2}\sum_{<i,j>} |w(r_i)-w(r_j)|^2} \quad (12)$$

where $\kappa$ and K are constants, the sum over $L$ means a sum over triangular lattices, and the indices i and j refer to orientation sensitive neurons located at the centers of the triangles in this lattice (note that N is the number of faces of the triangulation $L$ ). For large numbers of faces the triangulation $L$ can be thought of as approximating a smooth 2-dimensional surfaces, and in the limit N $\rightarrow \infty$ the sum over triangular lattices in eq.12 becomes a sum over smooth 2-dimensional surfaces. In this limit the partition function (12) becomes

$$Z = \int Dw(\sigma) \exp(-S)$$
, (13)

where $(\sigma_1, \sigma_2)$ are the coordinates of a point on the smooth surface and the continuum action S is given by

$$S = \frac{K}{2} \int d^2\sigma \partial_\alpha w \partial_\alpha w + \lambda$$
. (14)

The constant $\lambda$ in (14) replaces the constant $\kappa$ and plays the role of an energy per node. It turns out that the partition function (13) has an interesting physical interpretation [13]; namely, it represents the quantum theory of a "string" moving on a 2-dimensional surface - in mathematical terms this means random holomorphic mappings from a 2-dimensional manifold onto a fixed 2-dimensional manifold. In this string interpretation the angle variable w becomes a complex variable by the addition of a second real variable representing the local magnification of the mapping. It is worth noting that this result is consistent with the theorem [14] that for maps of 2-dimensional surfaces onto 2-dimensional surfaces the stationary state of Kohonen's algorithm is a holomorphic (or anti-holomorphic) map. Thus we have the general result that the feature vector will be a smooth analytic function of position, in accordance with our initial expectations. As noted in ref. 13 this formalism also determines the topological connectivity of the 2-dimensional surfaces involved; therefore in contrast with other approaches to information fusion the network topology is not an extra ad hoc assumption, but follows from the MDL principle.

We can now see why somewhat miraculously Kohonen self-organization provides an exact solution to the problem of finding minimally complex explanations for the outputs of a large number of sensors. In the large N limit explanations are represented by smooth mappings of a 2-dimensional surface representing the physical layout and connectivity of the network into feature space. The information cost of any particular explanation is just an exponential of minus the quantized area of the surface in feature space given in eq. 14. The natural unit of quantization, i.e. the area equivalent to 1 bit, is determined by the inverse of the constant K in eq. 14. The cost averaged over environmental inputs is just the negative logarithm of the partition function Z defined in eq. 13.

**4. Hierarchical versus distributed information fusion**

It is self-evident that other things being equal generation of a minimal binary representation for feature vectors and explanations for feature vectors will minimize the energy usage in any sensor network. A remaining question though is how to compare the energy costs of hierarchical Bayesian network with those of network that fuses sensor outputs via Kohonen self-organization. In a self-organized network of sensors and data processors the information fusion processes are distributed throughout the network. However if, as we have been implicitly assuming, the different sensors in the network are physically separated then some means must be provided for these nodes to communicate with each other. In a Bayesian network the communication support must be capable of relaying the conditional probabilities at one decision level of the network to the data fusion units in the next level of the network within a relevancy time interval. Thus an interesting question is how the data processing and communication energy costs for information fusion in a network using distributed self-organization compare with these costs in a network using hierarchical Bayesian reasoning.

If one uses a conventional hierarchical data fusion strategy [see e.g. 15] where separate data fusion nodes collect information from sensor nodes, then every data fusion node in the system must incorporate a Bayesian inference engine which calculates conditional probabilities for all the relevant hypotheses. In a Bayesian tree-like network of data fusion and sensor nodes with a total of N nodes, these conditional probabilities must be calculated at each of the N nodes and communicated to the node in the next higher level. If every data fusion node receives information from say 3 nodes in the next layer down, there will be approximately lnN layers and 2N/3 communication links in the network for large N. The total amount of information that needs to be transmitted from one layer of the network to the next will be on the order of (N/ lnN) $\sum_{\alpha}$ ($E_\alpha$ +(-ln $P(\alpha)$)) where $E_\alpha$ is the description cost for hypothesis α. On the other hand, in a network of M sensors using a self-organization scheme of the type discussed in the previous section for data fusion, the conditional probabilities are not directly calculated; instead they are coded into the description length for the feature vectors. This is a tremendous advantage because initially one can choose the most likely feature vector for every sensor node. Furthermore, after a certain number of iterations of Kohonen's algorithm the M feature vectors are compressed into a smooth function. Therefore in a self-organizing network the amount of information that must be processed during the data fusion process is enormously reduced because one isn't carrying along conditional probabilities for every possible hypothesis.

If every sensor node in a self-organizing network communicates with 3 of is neighbors, the number of communication links that must be established to implement Kohonen self-organization will be approximately equal to 3M/2. If we assume every data fusion node in a hierarchical Bayesian network also functions as a sensor node so N=M, we see that the minimum number of communication links required in a self-organizing network with the same number of sensors will be approximately 9/4 the required number of links in a hierarchical Bayesian network. However as one moves from one step of the data fusion process to the next the amount of information that must be transmitted in a self-organizing network will be vastly smaller when the number of hypotheses to be tested is very large. As discussed in the last section one can initiate the self-organization process by independently choosing feature vectors for each sensor according to the probability that they occur in the environment. However in reality the sensor readings are not independent, and it would make more sense to initially replace each sensor output with the most likely explanation for the sensor output. In this case the total amount of information transmitted between sensor nodes for each iteration of the Kohonen algorithm will be on the order of (3M/2) $F(x)$, where $F(x)$ is the average description cost for the explanations.

Since $F(x)$ will be on the order of $H E_\alpha$, where $H$ is the number of hypotheses, we see that the communication costs in a hierarchical Bayesian network will be larger than those in a self-organizing network with the same number of sensors by a factor on the order of $H$. If we assume that each data fusion node in a hierarchical Bayesian network combines the conditional probabilities from three nodes then the computational costs for each hypothesis are similar in the Bayesian and self-organizing networks. However in the Bayesian network the conditional probabilities must be updated for each hypothesis; therefore the computational cost per node will be on the order of $H$ larger in the Bayesian network. The end result is that the relative energy costs of moving from one step of the data fusion process to the next in a distributed self-organizing versus a hierarchical Bayesian network will be on the order of

$$\text{Bayesian energy cost / Self-organization energy cost} \sim \frac{H}{\ln N} \qquad (15)$$

## 5. Conclusion

We see that when there is only one hypothesis to test the energy cost for fusing sensor outputs using a hierarchical Bayesian network is not significantly different form the MDL energy costs of a self-organizing network. However, when the number of hypotheses to be tested is very large, then the energy costs of using distributed self-organization to fuse sensor outputs will be very much smaller. We should also note that because the fusion process in a self-organizing network is distributed throughout the network, using self-organization for information fusion also has the advantage of greater reliability. In addition we have seen that distributed self-organization may be better able to deal with the combinatorial complexities associated with ambiguous explanations. It is of course tempting to speculate that the energy saving and reliability features of distributed self-organization, as well as the ability to cope with ambiguous environmental stimuli, are principal reasons why biological evolution has favored self-organization and complete decentralization of the cognitive processes in the mammalian brain.

Acknowledgment: The author is grateful for discussions with Bob Bryant and Chris Cunningham.

## References
1. 8. J. Pearl, *Probabilistic Inference in Intellegent Systems* (Morgan Kaufmann 1988).
2. G. Rogova and R. Menon, "Decision Fusion for Learning in Pattern Recognition", in *Proceedings of the First International Conference on Multisource-Multisensor Fusion* (CREA Press Athens, Georgia 1998).
3. J. Rissanen, *Stochastic Complexity in Statistical Inquiry* (World Scientific 1998).
4. G. E. Hinton, et al , Science $\underline{268}$, 1158 (1995).
5. B. D. Ripley, *Pattern Recognition and Neural Networks* (Cambridge U. Press, 1996).
6. G. E. Hinton and T. J. Sejnowski, Cognitive Science $\underline{9}$, 147 (1985).
7. R. M. Neal, Neural Comp. $\underline{4}$, 832 (1992).
8. P. Dayan, G. Hinton, and R. Neal, Neural Comp. $\underline{7}$, 889 (1995).
9. W. A. Little, Math. Biosci. $\underline{19}$, 101 (1974); G. L. Shaw and R. Vasudevan, Math. Biosci. $\underline{21}$, 207 (1974).
10. G. Chapline, "Sentient networks", in *Proceedings of the First International Conference on Multisource-Multisensor Fusion* (CREA Press Athens, Georgia 1998).
11. T. Kohonen, *Self-Organization and Associative Memory* (Springer-Verlag 1987).
12. Ritter, H. and Schulten, K. , *IEEE International Conference on Neural Networks* $\underline{I}$ 109 (1988).
13. G. Chapline, Network: Comp. Neural Syst. $\underline{8}$, 185 (1997).
14. Ritter, H. and Schulten, K., *Biol. Cybern.* $\underline{54}$ 99 (1986).
15. R. Lobbia and M. Rakijas, SPIE $\underline{3081}$ , 1$\underline{10}$ (1997).